\documentclass[aps,prl,superscriptaddress,reprint]{revtex4-1}
\usepackage{amsmath}
\usepackage{amssymb}
\usepackage{graphicx}
\usepackage{hyperref}
\usepackage{url}
\usepackage{color,xcolor}
\usepackage{ulem}
\begin{document}

\title{Sequential structural and antiferromagnetic transitions in BaFe$_2$Se$_3$ under pressure}
\author{Yang Zhang}
\author{Ling-Fang Lin}
\author{Jun-Jie Zhang}
\affiliation{School of Physics, Southeast University, Nanjing 211189, China}
\author{Elbio Dagotto}
\affiliation{Department of Physics and Astronomy, University of Tennessee, Knoxville, TN 37996, USA}
\affiliation{Materials Science and Technology Division, Oak Ridge National Laboratory, Oak Ridge, TN 37831, USA}
\author{Shuai Dong}
\email{Corresponding author. E-mail: sdong@seu.edu.cn}
\affiliation{School of Physics, Southeast University, Nanjing 211189, China}
\date{\today}

\begin{abstract}
The discovery of superconductivity in the two-leg ladder compound BaFe$_2$S$_3$ has established the $123$-type iron chalcogenides as a novel and interesting subgroup of the iron-based superconductors family. However, in this 123 series, BaFe$_2$Se$_3$ is an exceptional member, with a magnetic order and
crystalline structure different from all others. Recently, an exciting experiment reported the emergence of superconductivity in BaFe$_2$Se$_3$ at high pressure [J.-J. Ying, H. C. Lei, C. Petrovic, Y.-M. Xiao, and V.-V. Struzhkin, Phys. Rev. B \textbf{95}, 241109 (\textbf{R}) (2017)]. In this publication, we report a first principles study of BaFe$_2$Se$_3$. Our analysis unveils a variety of qualitative differences between BaFe$_2$S$_3$ and BaFe$_2$Se$_3$, including in the latter an unexpected chain
of transitions with increasing pressure. First, by gradually reducing the tilting angle of iron
ladders, the crystalline structure smoothly transforms from $Pnma$ to $Cmcm$ at $\sim 6$~GPa. Second,
the system becomes  metallic at $10.4$~GPa. Third, its unique ambient pressure Block antiferromagnetic ground
state is replaced by the more common CX antiferromagnetic order at $\sim 12$~GPa, the same magnetic state of
the 123-S ladder. This transition is found at a pressure very similar to the experimental superconducting transition.
Finally, all magnetic moments vanish at $30$~GPa. This reported
theoretical diagram of the complete phase evolution is important because of the technical challenges to capture many
physical properties in high-pressure experiments. The information obtained in our calculations
suggest different characteristics for superconductivity in BaFe$_2$Se$_3$ and BaFe$_2$S$_3$: in 123-S pairing
occurs when magnetic moments vanish, while in 123-Se the transition region from Block- to CX-type magnetism appears
to catalyze superconductivity. Finally, an additional superconducting dome above $\sim 30$~GPa is expected to occur.
\end{abstract}

\maketitle

\section{Introduction}
Since the initial discovery of superconductivity in the iron pnictides, the study of
iron chalcogenides have rapidly developed into another extensively addressed branch of iron-based
superconductors~\cite{Dai:Np,Stewart:Rmp,Dagotto:Rmp,Dai:Rmp}. For the vast majority of these
novel superconductors, the iron lattice of relevance is quite similar: a
slightly distorted two-dimensional iron square lattice stacking along the
$c$-axis, where each iron atom is caged in a tetrahedral structure coordinated by pnictogens or
chalcogens~\cite{Mazin:np,Johnston:Ap,Dagotto:Rmp,Dai:Rmp,Si:NRM}. Different from the iron pnictides,
the iron chalcogenides usually display a larger local magnetic
moment~\cite{Li:Np,WangMeng:Prb,Dagotto:Rmp,zhang:prm,Bao:Cpl} and Fermi surface without
hole pockets, implying that the physical mechanism for superconductivity cannot be simply based
on Fermi surface nesting considerations in the weak coupling Hubbard $U$ limit~\cite{Dai:Np,Dagotto:Rmp}.

\begin{figure}
\centering
\includegraphics[width=0.48\textwidth]{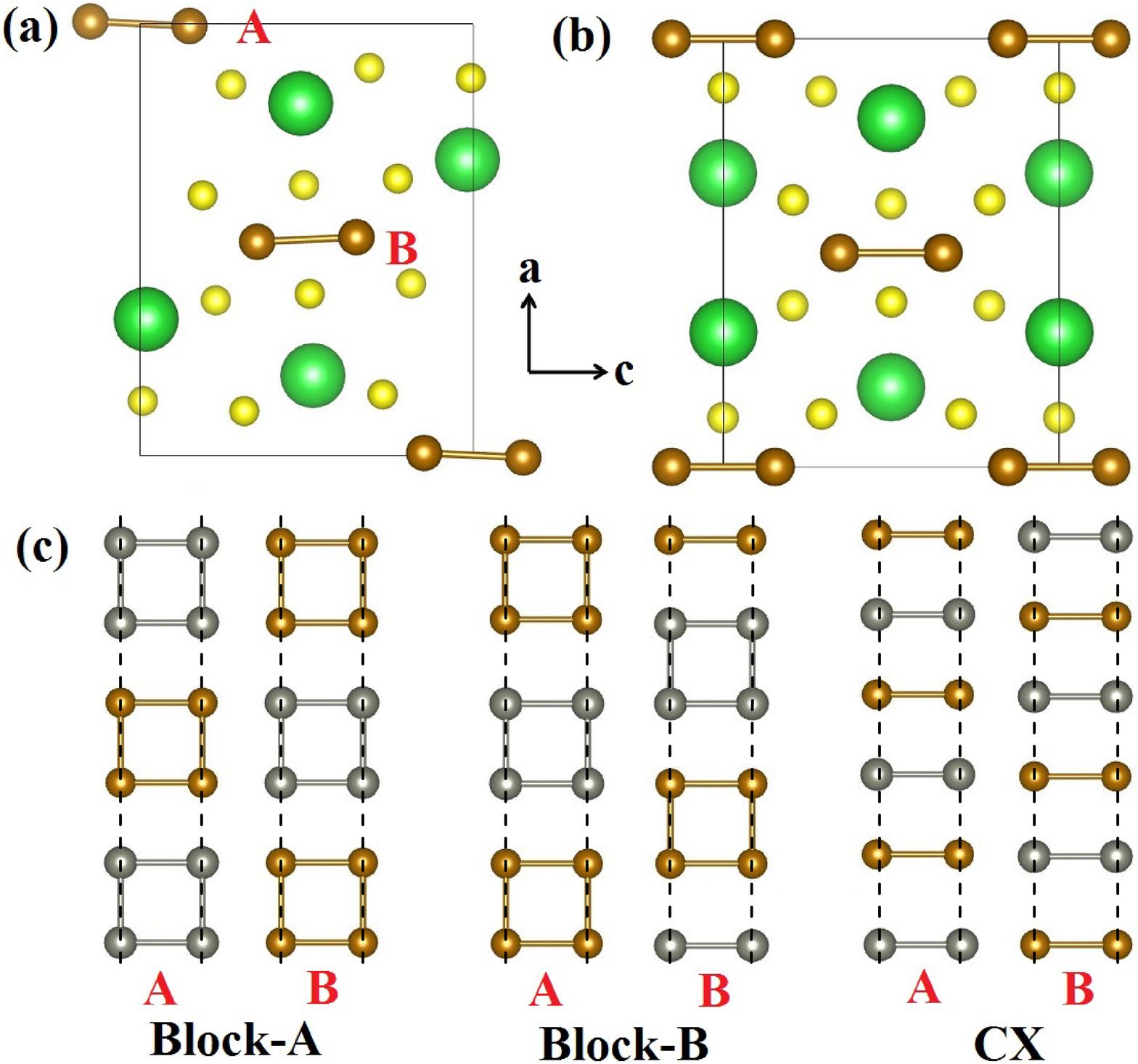}
\caption{(a-b) Schematic crystal structures of $A$Fe$_2X_3$ with the convention:
Green = $A$; Yellow = $X$; Brown = Fe. (a) corresponds to the space group No. 62 $Pnma$.
(b) corresponds to the space group No. 63 $Cmcm$. The key difference between these structures
is the tilting/non-titting of the iron ladders. (c) Sketch of some possible spin patterns studied here.
Spin up and down are distinguished by colors brown and silver, respectively.
A and B are the ladder indexes in one unit cell, as indicated in (a-b). Note A and B are ladders
located in different layers.}
\label{Fig1}
\end{figure}

Recently, the two-ladder iron chalcogenides, with the $123$-type $A$Fe$_2X_3$ ($A$=K, Cs, Rb,
or Ba; and $X$=S, Se, or Te) chemical composition, have received considerable attention
due to their interesting physical properties and unique quasi-one-dimensional
structure~\cite{Maziopa:Jpcm,Nambu:Prb,Saparov:Prb,Lei:Prb,du:prb12,luo:prb13,monney:prb13,hirata:prb15,mourigal:prl15,Arita:prb,Wang:prb16,chi:prl,Zhang:prb17,Patel:prb,Patel:prb17}
(see Fig.~\ref{Fig1}). Remarkably, it has been found experimentally
that BaFe$_2$S$_3$ is the first iron-ladder that becomes superconducting,
under pressures above $10$~GPa and with a $T_{\rm c}=24$~K~\cite{Takahashi:Nm,Yamauchi:prl15}. At ambient
conditions, BaFe$_2$S$_3$ is a Mott insulator with the so-called CX-type antiferromagnetic order
[see Fig.~\ref{Fig1}(c), with antiferromagnetic legs and ferromagnetic rungs]
below $120$ K~\cite{Takahashi:Nm}. This CX order is sometimes referred to as $(\pi,0)$ order as well.
Our recent calculation based on
density functional theory (DFT) addressed the evolution of the magnetic/electronic properties of
BaFe$_2$S$_3$ under pressure~\cite{Zhang:prb17}. With increasing pressure, we found that the magnetic moments
in BaFe$_2$S$_3$ abruptly reduce to zero at a critical pressure. The Mott gap closes slightly
in advance, i.e. at a smaller pressure, leading to a non-magnetic (NM) metallic phase, presumably with
short-range CX magnetic order,
which may be the prerequisite for superconductivity~\cite{Zhang:prb17}.
Similar transitions were also predicted for other members of the
123 series, such as KFe$_2$S$_3$~\cite{Zhang:prb17}.

BaFe$_2$Se$_3$ is another important member of the iron
two-leg ladder family. This material is an antiferromagnetic (AFM) Mott insulator
with a robust N\'eel temperature ($T_{\rm N}$) $\sim 256$~K~\cite{Caron:Prb12}. More importantly, BaFe$_2$Se$_3$
is an exceptional member of the 123 series because its physical properties are qualitatively different from
others. First, although the members of the 123 series tend to form an orthorhombic structure, the space group of
BaFe$_2$Se$_3$ is fairly unique, namely the $Pnma$~\cite{Caron:Prb}, while all other iron ladders
share the more common $Cmcm$
space group~\cite{Takahashi:Nm,Caron:Prb12} [see sketches in Figs.~\ref{Fig1}(a-b)]. The most
clear distinction  related with the different space groups is the tilting of the ladders that occurs
in the 123-Se case. Second, the magnetic ground
states are totally different between the BaFe$_2$Se$_3$ and other 123 ladders. More specifically, BaFe$_2$Se$_3$
hosts an exotic block-type AFM order with a large magnetic moment ($2.8$ $\mu_{\rm B}$/Fe), as opposed to a CX AFM state.
It should be remarked that the block-type AFM order is rare in iron pnictides/chalcogenides,
appearing only in a few materials~\cite{Li:Np,zhang:rrl,Ye:Prl}. By contrast, all other 123 ladders
host the more conventional CX-type AFM order, with smaller magnetic
moments~\cite{Caron:Prb,Wang:prb16,Wang:prb17}. These two differences are highly nontrivial.
For example, the block-type AFM order can drive improper ferroelectricity in the $Pnma$ structure,
rendering BaFe$_2$Se$_3$ a potential high temperature multiferroic material~\cite{Dong:PRL14,Lovesey:PS}.

Considering its robust magnetic characteristics (large local moments and high $T_{\rm N}$ for the block
AFM state)~\cite{Caron:Prb12} and its large band gap ($>0.5$ eV from DFT calculation)~\cite{Dong:PRL14},
BaFe$_2$Se$_3$ seems to be far from superconductivity, according to the empirical knowledge gathered on iron-based
superconductors. However, a striking experimental discovery was recently reported in this compound:
superconductivity can also be induced in BaFe$_2$Se$_3$
under high  pressure, in the range $10.2$-$15$~GPa~\cite{Ying:prb17}.
The highest superconducting $T_{\rm c}$ reaches $\sim 10$~K.
Interestingly, the local magnetic moments remain considerable large in
BaFe$_2$Se$_3$ under high pressure, as shown using the integrated absolute difference (IAD)
analysis. These  local magnetic moments reduce to zero only when the pressure reaches
$30$~GPa. This overall phenomenology is nontrivial, because it
means that the superconducting phase emerges directly from
a magnetic phase with a large spin moment ($>1.3$ $\mu_{\rm B}$/Fe). In contrast, for
BaFe$_2$S$_3$ the system is already in a non-magnetic metallic state
when pressure is larger than $10.8$~GPa,
according to DFT calculations~\cite{Zhang:prb17}. This result is in close proximity to the experimental
critical pressure for superconductivity ($\sim 11$~GPa)~\cite{Takahashi:Nm}. Therefore,
the underlying physical mechanisms regarding the superconducting pairing may have different characteristics
between BaFe$_2$S$_3$ and BaFe$_2$Se$_3$, which is intriguing.
Possible structural and magnetic transitions
were also discussed in Ref.~\cite{Ying:prb17}. However, due to the technical challenges intrinsic of high
pressure experiments, the available experimental information regarding the evolution of
crystalline/electronic/magnetic structures remains limited, which prevents a complete
understanding of BaFe$_2$S$_3$ under pressure. Theory is needed to guide the physical description
of this compound.

In this publication, a systematic study of the physical evolution of BaFe$_2$Se$_3$ under pressure is reported
using first-principles DFT calculations. Contrary to the straightforward CX-AFM insulator to NM metal
transition observed in BaFe$_2$S$_3$ with increasing pressure, here a far
more complex evolution involving four transitions has been found for BaFe$_2$Se$_3$.

%Four transitions have been  First, similar results as in previous DFT efforts and experiments for the ambient condition have been obtained. Second, we find the structural transition from $Pnma$ to $Cmcm$ at $6$ GPafrom $Pnma$ to $Cmcm$ at $6$ GPa. Third, a insulator-metal transition and AFM transition from Block-type to CX-type have been obtained under pressure. Our results also find that metallic bond behavior and electron transfer from Se to Fe may be play a critical role to superconductivity, where superconductivity may cause by both fluctuation of charge and spin.

\section{Method}
The DFT calculations were performed using the Vienna {\it ab initio} simulation
package (VASP), with the projector augmented-wave (PAW)
potentials~\cite{Kresse:Prb,Blochl:Prb,Perdew:Prl}. The Perdew-Burke-Ernzerhof (PBE) exchange
function was employed~\cite{Kresse:Prb96} and the plane-wave cutoff energy
was $500$ eV. For the block-type AFM order a $3\times3\times4$ $k$-point mesh was used.
Since the minimal unit cell is different for different magnetic states,
this mesh was appropriately modified for the other various magnetic cells considered
(e.g. $3\times7\times2$ for the CX-AFM order) to render the $k$-point densities approximately
the same in reciprocal space. In addition, we have tested that these $k$-point meshes already
lead to converged energies when compared with denser meshes. Both the lattice constants
and atomic positions were fully relaxed until the force on each atom was below $0.01$ eV/{\AA}.

To study the magnetic properties, various possible (in-ladder) magnetic arrangements
were imposed on the iron ladders, such as NM, ferromagnetic (FM), CX-AFM,
CY-AFM (AFM rungs and FM legs), G-AFM (both rungs and legs AFM), and block-AFM~\cite{Zhang:prb17}. Despite the dominance of
the in-ladder magnetic order, the magnetic correlations between ladders
can also slightly affect the energies and physical properties.
Therefore, the ($\pi$, $\pi$, $0$) order was adopted for the CX-AFM arrangement,
as suggested by neutron scattering results and by our previous DFT
investigations~\cite{Dong:PRL14,Zhang:prb17}. Similarly, two types of block-AFM
magnetic patterns (Block-A and Block-B) were also considered~\cite{Dong:PRL14}.

In our calculations, starting from the experimental lattice constants under ambient conditions~\cite{Caron:Prb},
the structures with increasing pressures are obtained via structural optimization in the DFT calculation,
since no experimental structural data at different pressures are available thus far.

\section{Results}

\subsection{Physical properties under ambient conditions}

Before describing the calculations with pressure, the basic DFT results
corresponding to BaFe$_2$Se$_3$ without external pressure are
briefly reviewed here. These results were previously
reported  by some of the authors~\cite{Dong:PRL14}. In those investigations,
it was observed that the pure GGA approach is the best to describe BaFe$_2$Se$_3$,
consistent with previous DFT calculations of BaFe$_2$S$_3$~\cite{Suzuki:prb,Zhang:prb17}.
Thus, the GGA exchange will be adopted for all the following calculations.
Our main previous DFT results are summarized in Table~\ref{Table1} for the benefit of the readers.

First, under ambient conditions, the Block-B magnetic order has the lowest energy
among all tested magnetic orders, in agreement with neutron
experiments~\cite{Caron:Prb12}. For the Block-B state, the calculated local
magnetic moment of Fe is about $2.88$ $\mu_{\rm B}$/Fe, quite close to the
experimental value~\cite{Caron:Prb}. This moment is much higher than the
magnetic moment of BaFe$_2$S$_3$ ($\sim 1.2$ $\mu_{\rm B}$/Fe)~\cite{Takahashi:Nm}.

Second, the calculated lattice constants also agree well with the experimental
values \footnote{The small discrepancy of lattice constants between theory and experiments along
the $a$-axis is probably due to the complex magnetic coupling between ladders
along that $a$-axis, as discussed in Ref.~\cite{Lovesey:PS}. However, a very large
cluster is required to accommodate such a complex magnetic coupling,
beyond the goals of the present study.}. For the Block-B type phase, the Fe-Fe distances
are dimerized: $2.82$ {\AA} and $2.58$ {\AA} due to
magnetostriction effects~\cite{Dong:PRL14}, also in agreement with neutron
experiments~\cite{Caron:Prb12}.

Third, the energy gap corresponding to the Block-B AFM order is about $0.5$ eV, in agreement
with previous DFT calculations~\cite{Saparov:Prb,Dong:PRL14}. Although Ref.~\cite{Lei:Prb}
reported a much smaller gap obtained from transport curves, this discrepancy is probably
caused by non-stoichiometry, a quite common phenomenon in BaFe$_2$Se$_3$. The optical
gap is needed to find the intrinsic band gap.

Fourth, according to the calculated density of states of the Block-B AFM state
(not shown here), the bands near the Fermi level are primarily made of
Fe-$3d$ orbitals hybridized with Se-$4p$ orbitals, and moreover the Fe atoms are in the
high spin state.

\begin{table}
\centering
\caption{The optimized
local magnetic moment (in $\mu_{\rm B}$/Fe unit) within the default PAW sphere,
lattice constants ({\AA}),
band gaps (eV) for the various magnetic structures,
and energy differences (meV/Fe) with respect to the NM configuration taken as the
reference of energy.
The experimental values (Exp. for short) are also listed for comparison.}
\begin{tabular*}{0.48\textwidth}{@{\extracolsep{\fill}}llllc}
\hline
\hline
  & $M$ & $a$/$b$/$c$ & Gap  & Energy \\
\hline
NM       &   0  &11.23/5.38/9.09 & 0   &    0    \\
FM      & 2.96 &12.07/5.41/9.2  & 0  & -105.9    \\
CX      & 2.46 &11.94/5.41/9.11 & 0.11  & -261.1 \\
CY       & 2.82 &12.12/5.42/9.24 & 0.15  & -221  \\
Block-A  & 2.89 &12.14/5.40/10.65& 0.46 & -274.8 \\
Block-B & 2.88 &12.15/5.40/9.17 & 0.50   & -282.7 \\
Exp.\cite{Nambu:Prb} &-- & 11.93/5.44/9.16  & --&--\\
Exp.\cite{Caron:Prb} & 2.8   & 11.88/5.41/9.14 & --&--\\
Exp.\cite{Lei:Prb} & --  &--  & 0.178 &--\\
\hline
\hline
\end{tabular*}
\label{Table1}
\end{table}

\subsection{Transitions under pressure}

\begin{figure}
\centering
\includegraphics[width=0.48\textwidth]{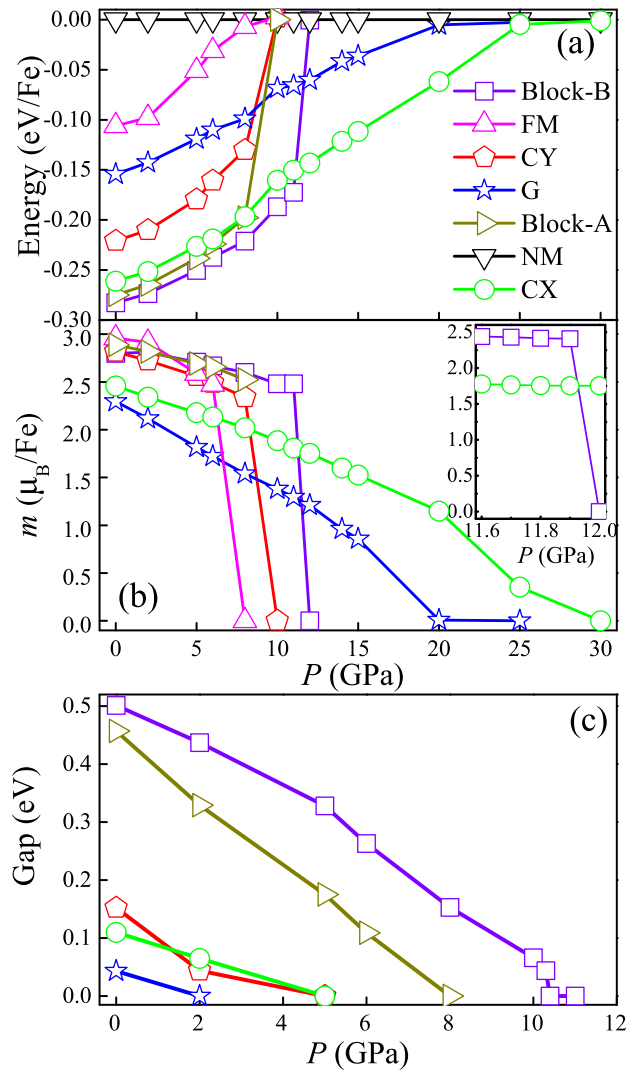}
\caption{Evolution of the magnetic and electronic structures of BaFe$_2$Se$_3$ as a function
of pressure. (a) Energies (per Fe) of the various magnetic orders indicated. (b) Local magnetic
moments of Fe, integrated within the default Wigner-Seitz sphere as specified by VASP.
{\it Inset:} an amplified view near the transition. (c) Band gaps for the many states analyzed.}
\label{Fig2}
\end{figure}

Consider now the effect of hydrostatic pressure when introduced in the calculation. Both the
lattice constants as well as the atomic positions are fully relaxed again~\cite{Zhang:prb17}.
The calculated energies for the various magnetic states considered here
are shown in Fig.~\ref{Fig2}(a),
as a function of pressure. The Block-B order has the lowest energy until approximately $12$~GPa.
In that range of pressures, the CX state is slightly higher in energy than Block-A and Block-B, but much lower
than others. The crossover between Block-B and CX occurs at $\sim 12$~GPa, suggesting a
pressure induced magnetic phase transition. In fact at $\sim 12$~GPa, the energy of the Block-B magnetic arrangement
displays a sudden jump, and the state becomes degenerate with the NM state. Starting at $\sim 12$~GPa,
the CX state holds the lowest energy until $30$~GPa,
where all the magnetic states give the same identical energies as the non-magnetic one.

In principle, the enthalpy should be used to determine the phase transition when dealing
with the condition of fixed pressure and varying volume, which is common in the context of pressure-induced
structural transitions. However, the phase transitions involved here are mainly magnetic-related, instead of
structural. The volume difference between the Block-B state and CX state is within $2\%$ in the range of 0-10 GPa.
In the experiment (Ref.~\cite{Ying:prb17}), the pressure was applied to the sample using diamond anvils at room
temperature. Then the sample was cooled down at a fixed volume (fixed by diamond anvils and surrounding wrap).
The pressure value at room temperature is used as the experimental value. Strictly speaking, the real experimental
conditions were not fixed-pressure, but more likely fixed-volume, which are different from other
high pressure experiments on structural transitions. Thus, the most precise theoretical treatment should be carried out
by comparing the total energies of different magnetic orders using the experimental lattice constants, as done
in Ref.~\cite{Aoyama:Nc} for pressured TbMnO$_3$. However, since the experimental structural information of
BaFe$_2$Se$_3$ under pressure is unavailable at the current stage, the next-best choice is to use the optimized
structure for each magnetic order, as done in our work \footnote{We also tested the use of the enthalpy
to determine the phase transitions. Unfortunately, using calculations based on enthalpy, a premature
phase transition from Block-B to CX order occurs at $\sim3$ GPa, together with the insulator-metal
transition, which is very different from the available experimental information (insulator-metal transition
occurs at $10$ GPa). Although a fine tuning of the Hubbard $U$ in the DFT+$U$ formalism can be used so that
the insulator-metal transition matches the experimental value, the side effect is that there is a robust
magnetic state at $30$ GPa, which is different from the experimental information. In summary, the use of
the enthalpy leads to large discrepancies between the theoretical and experimental phase diagrams even if $U$ is tuned.}.

With increasing pressure from zero, the local magnetic moment of the iron atoms decreases monotonously,
as shown in Fig.~\ref{Fig2}(b), for all the proposed states. For the interesting Block-B state that is
the ground state at ambient conditions, the moment
slowly decreases from $2.88$ $\mu_{\rm B}$/Fe at $0$~GPa to $2.41$ $\mu_{\rm B}$/Fe
at $11.9$~GPa, and then abruptly drops to zero as in a first-order transition,
which corresponds also to the sudden jump in energy shown in Fig.~\ref{Fig2}(a) for this state. By
contrast, in the case of the CX phase the local magnetic moment decreases to zero continuously
until $30$~GPa, resembling a second-order transition. Considering the magnetic ground
state transition from Block-B to CX at $\sim 12$~GPa [Fig.~\ref{Fig2}(a)], the local magnetic
moment should persist to be nonzero in the range from $0$~GPa to $30$~GPa, which agrees
with the experimental observation~\cite{Ying:prb17} (in addition, we predict that
a sudden reduction in the magnetic moment should be observed experimentally
at $\sim 12$~GPa).

Between approximately $10$ and $15$~GPa (the region where the superconducting
dome was found experimentally), the calculated local magnetic moment
is about $1.75$-$1.53$ $\mu_{\rm B}$/Fe for the
normal state. This is very close, and only slightly higher, than the experimental value obtained
via the IAD analysis at $17$~K~\cite{Ying:prb17}. It should be noticed that finding overestimated
magnetic moments is quite common in DFT calculations of stripe
AFM order (C-type) in iron-based superconductors ~\cite{Hansmann:prl10,Mazin:np,Yin:Nm10,Suzuki:prb},
which may due to coexisting localized Fe spins and itinerant electrons. In addition, the methods
to estimate the ``local" magnetic moments are different between DFT and neutron scattering. In the
latter, due to fluctuations, the time averaged magnetic moment is measured and usually this is smaller
than the actual instantaneous local moment.
This behavior with nonzero moments is conceptually different from that found in BaFe$_2$S$_3$, whose local magnetic moment quenches to zero in the normal
state corresponding to the superconducting dome region, according
to our previous DFT calculations~\cite{Zhang:prb17}.

The band gaps of various magnetic orders are displayed in Fig.~\ref{Fig2}(c). With increasing
pressure, all gaps decrease monotonously and eventually close. Considering the aforementioned magnetic
transition, the system persists to be insulating until the collapse of the Block-B state gap, beyond which
the system is metallic even still within the Block-B magnetic state. The critical pressure ($10.4$~GPa)
for this insulator-metal transition is also very similar  to the experimental observation
($10.2$~GPa for metallic behavior at low temperatures)~\cite{Ying:prb17}. After the magnetic order
of the Block-B phase collapses at $\sim 12$~GPa, then the ground state is metallic and has CX order.

\begin{figure}
\centering
\includegraphics[width=0.48\textwidth]{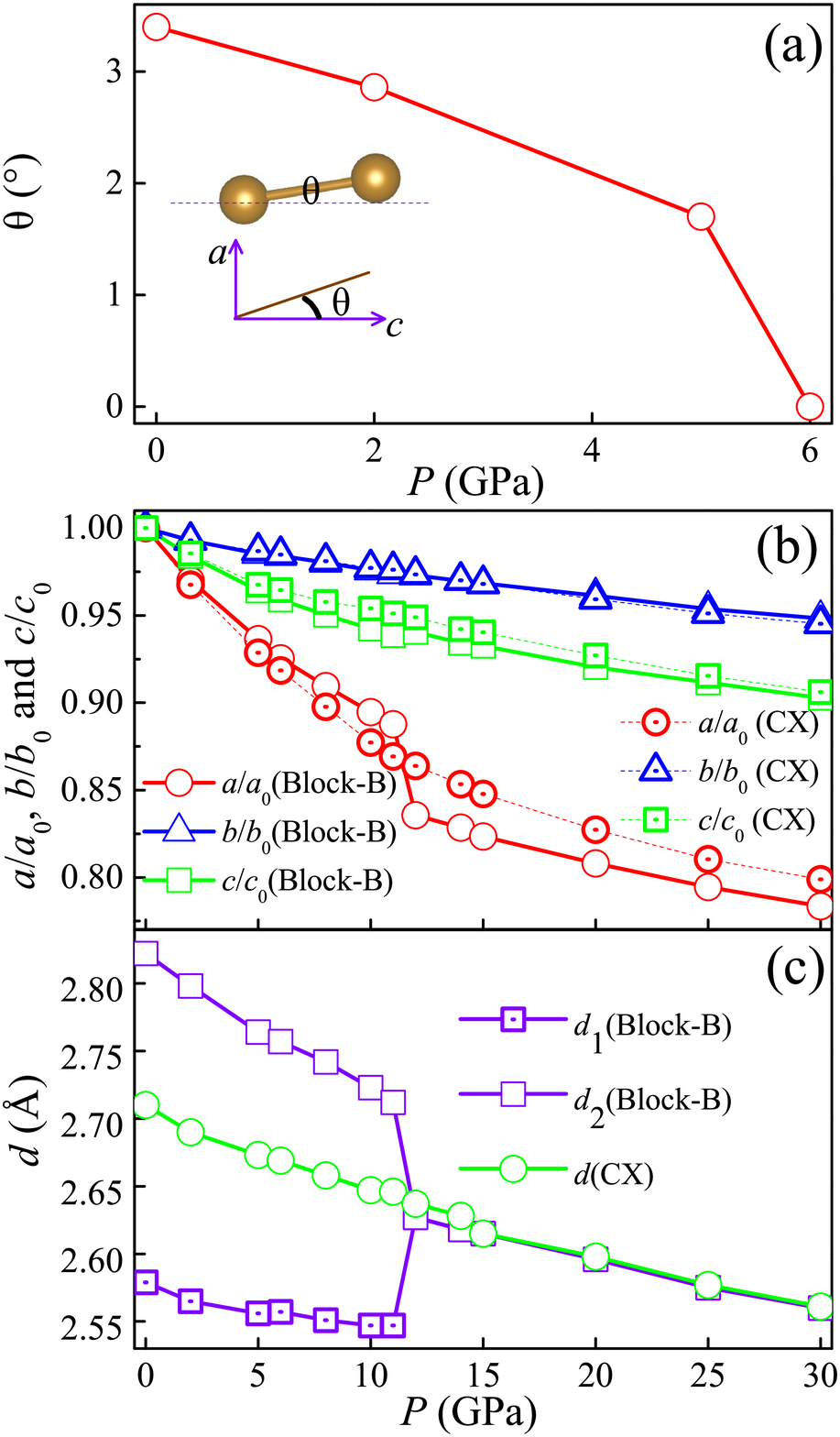}
\caption{Evolution of the crystalline structure of BaFe$_2$Se$_3$ varying pressure.
(a) The tilting angle of the iron ladders in the Block-B magnetic state
(see sketch in panel (a) of Fig.~1).
(b) Lattice constants, normalized to the original
ones at zero pressure.
(c) The Fe-Fe distance along the leg ladder direction. In both (b) and (c), the results for
the dominant CX and Block-B phases are shown for comparison. In (c), there are two distances
for the Block-B phase because there are two types of bonds along the legs: AFM and FM.}
\label{Fig3}
\end{figure}

In addition to the changes in the magnetic state and iron moments,
a structural evolution has also been observed, as shown in Fig.~\ref{Fig3}(a).
As schematically displayed in the inset, in Fig.~\ref{Fig3}(a) we show the
tilting angle of the  iron ladders, defined as $\theta$, with respect to the $b-c$ plane.
Its nonzero value is the main difference between the $Pnma$
(BaFe$_2$Se$_3$) and $Cmcm$ structures (other $A$Fe$_2X_3$'s). Under pressure, this tilting
angle gradually decreases to $0^{\circ}$ at $\sim 6$~GPa, implying a structural transition from
BaFe$_2$Se$_3$ ($Pnma$) into $\gamma$-BaFe$_2$Se$_3$ ($Cmcm$)~\cite{Svitlyk:JPCM}.

Figure~\ref{Fig3}(b) is a measure of the anisotropic compressibility, illustrating
that fact that the lattice is softest along the $a$-axis (perpendicular to the ladders)
and hardest along the $b$-axis (the leg direction of the ladders).
This is reasonable: the key substructures of the crystal are the two-leg ladders that
are difficult to modify with pressure. Such anisotropic compressibility is qualitatively similar
for all the 123 series, due to the spare space between ladders and the compact bonds along
ladders~\cite{Zhang:prb17}. The magnetic transition at $12$~GPa also leads to a discontinuous
change in the lattice constants, although the space group remains the same. The tetramerization
of the Block-B magnetic state, characterized by a disproportionation
of the Fe-Fe bond lengths, is suppressed by increasing
pressure from zero. At the critical pressure ($\sim 12$~GPa), the structural tetramerization suddenly
disappears, accompanying the magnetic transition, as summarized in Fig.~\ref{Fig3}(c).

\section{Discussion}

\begin{figure}
\centering
\includegraphics[width=0.48\textwidth]{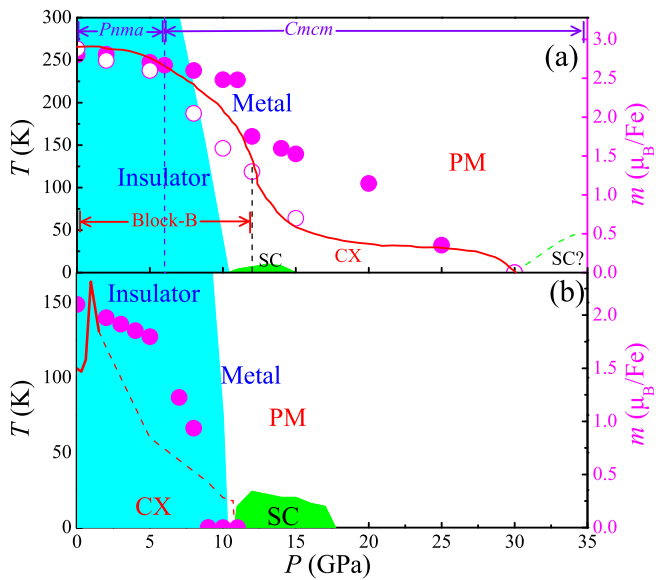}
\caption{Phase diagram of BaFe$_2X_3$ ($X$= S, Se) varying pressure, according to experimental
data \cite{Ying:prb17,Caron:Prb,Takahashi:Nm} and to our present and previous \cite{Zhang:prb17}
DFT calculations. The curves of AFM N\'eel temperatures (red solid), including the superconducting (SC) domes,
are all from experiments. For example, see Fig.~4 of Ref.~\cite{Ying:prb17}.
(a) Results for BaFe$_2$Se$_3$. The superconducting dome beyond $30$~GPa
is a theoretical prediction, to be verified in future experiments.
(b) Results for BaFe$_2$S$_3$. The dashed portion of the curve for
$T_{\rm N}$ is an assumed smooth drop with increasing pressure (experimental data is lacking).
Right axis: local magnetic moments. Open symbols: experimental data at low temperature.
Solid symbols: DFT data (at zero temperature).}
\label{Fig4}
\end{figure}

According to the results gathered in this study, it is possible to sketch a theoretical phase diagram for
BaFe$_2$Se$_3$ under pressure, as shown in Fig.~\ref{Fig4}(a). With increasing pressure, the
structural transition occurs first at $6$~GPa, with second order characteristics. The $Pnma$
group in the $A$Fe$_2X_3$ structure changes to the more
common $Cmcm$ group. No electronic/magnetic anomaly is visible
at this structural transition. As the pressure continues increasing,
an insulator-metal transition occurs at $10.4$~GPa, slightly before
the magnetic transition (from Block-B to CX states) that takes place at $12$~GPa. Remarkably, the experimentally
observed superconducting dome (pairing is beyond the capabilities of DFT)
emerges very close to the theoretical boundary between the Block-B and CX magnetic states.
Finally, the CX phase persists until a very high
pressure of approximately $30$~GPa where all magnetic moments vanish.
This phase diagram, including the evolution of the local magnetic moments,
either agrees well or is compatible with the available experimental data.
In addition, our results also provide additional systematic information, such as the unexpected
transition from Block-B to CX states. This transition is difficult to observe
experimentally because of the challenges in dealing with high pressure setups.

For comparison, the phase diagram of BaFe$_2$S$_3$ is shown in Fig.~\ref{Fig4}(b), based on data from
previous DFT~\cite{Zhang:prb17} and experimental efforts~\cite{Takahashi:Nm,chi:prl}. This phase
diagram is much simpler. In this case the system turns directly from an insulator with CX magnetic order
to a non-magnetic metal with increasing pressure, and
the experimental superconducting dome emerges at the CX-NM boundary. As a reasonable
conclusion of the S-based phase diagram, the driving force of superconductivity
may be attributed to short-range magnetic
fluctuations of CX-type in the non-magnetic state. However, according to DFT the driving force of superconductivity
in BaFe$_2$Se$_3$ appears to be induced, instead, by the competition between
Block-B and CX magnetic orders. This is an exotic scenario because it involves two states with long-range
magnetic order, and a first-order-like transition between them, contrary to the more established framework with
magnetic and nonmagnetic states as in BaFe$_2$S$_3$.
This difference may be the reason why the
highest superconducting $T_{\rm c}$ in the Se-ladder is considerably
lower than in the S-ladder.

In spite of these differences, it is interesting to note that in both materials
the local magnetic moments fast drop by $\sim1$ $\mu_{\rm B}$/Fe just before the appearance of the superconducting
dome. In this sense, the spin fluctuations are somewhat similar in BaFe$_2$S$_3$ and BaFe$_2$Se$_3$ at the
critical points. Of course, the physical mechanism(s) for superconductivity in these two system needs
further theoretical and experimental investigations.

In addition, according to the experience gathered on BaFe$_2$S$_3$,
an interesting speculation for BaFe$_2$Se$_3$ is that another superconducting dome, potentially
with a higher $T_{\rm C}$, may exist a little above $30$~GPa.

\begin{figure}
\centering
\includegraphics[width=0.48\textwidth]{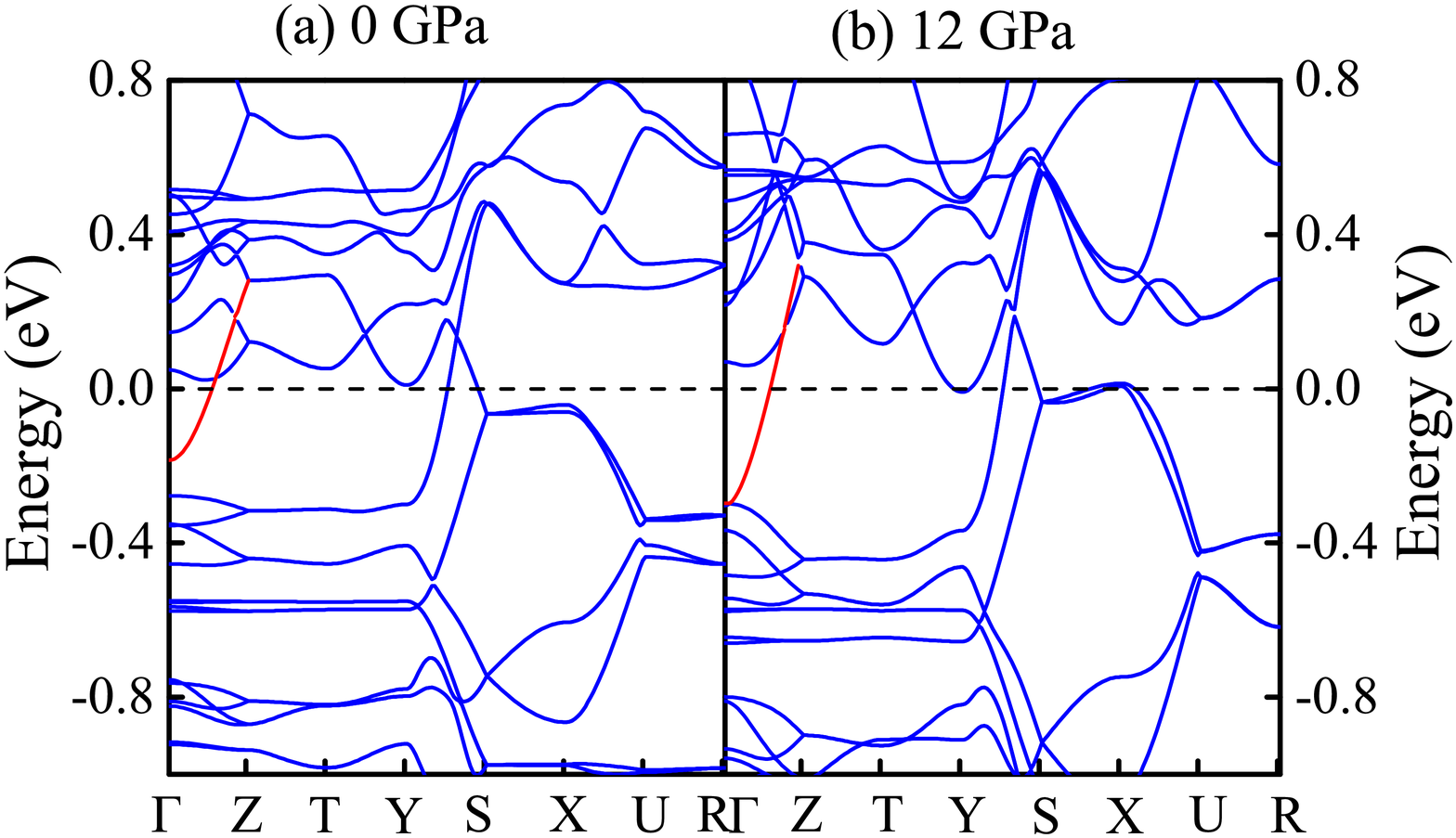}
\includegraphics[width=0.48\textwidth]{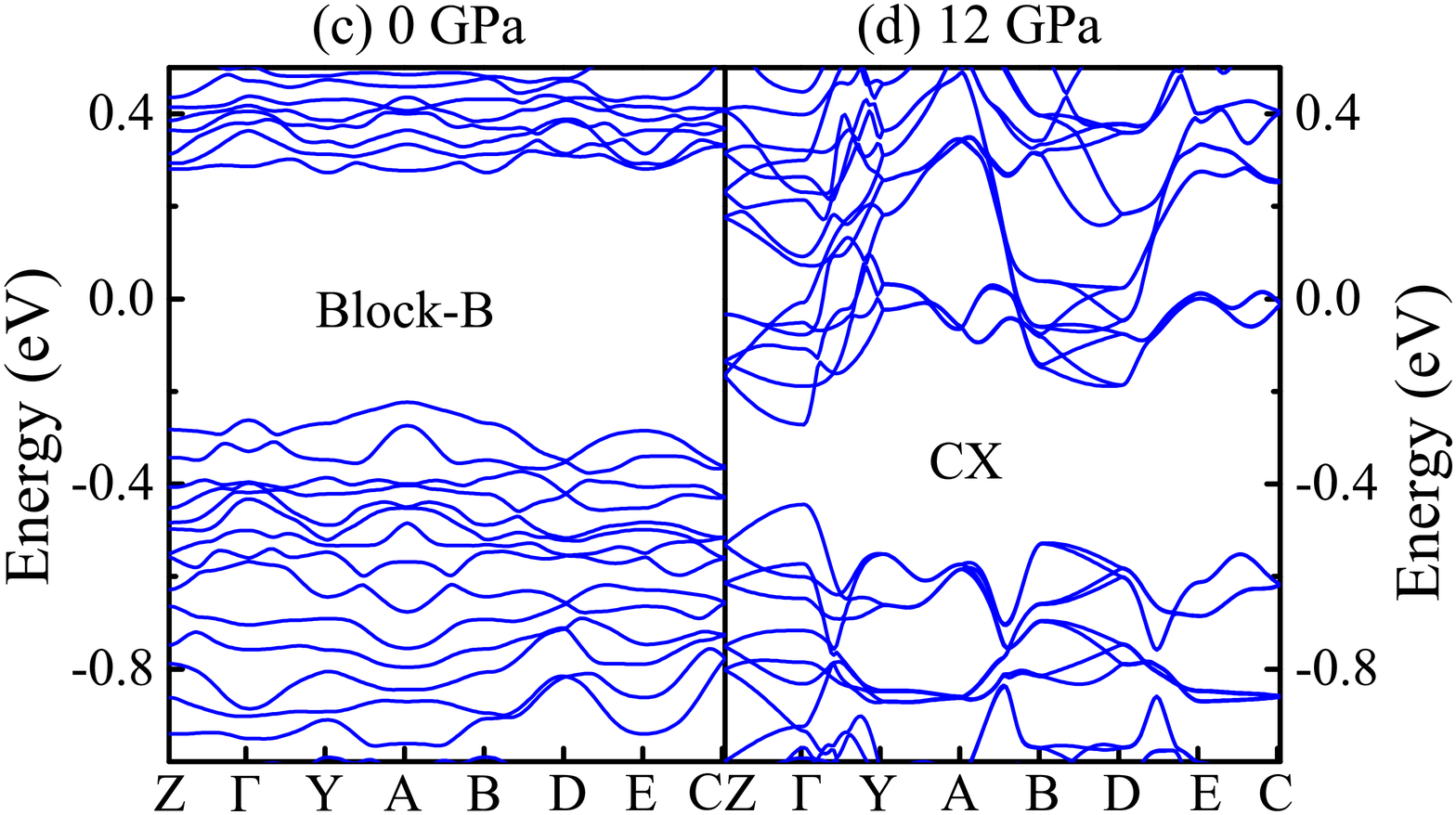}
\includegraphics[width=0.48\textwidth]{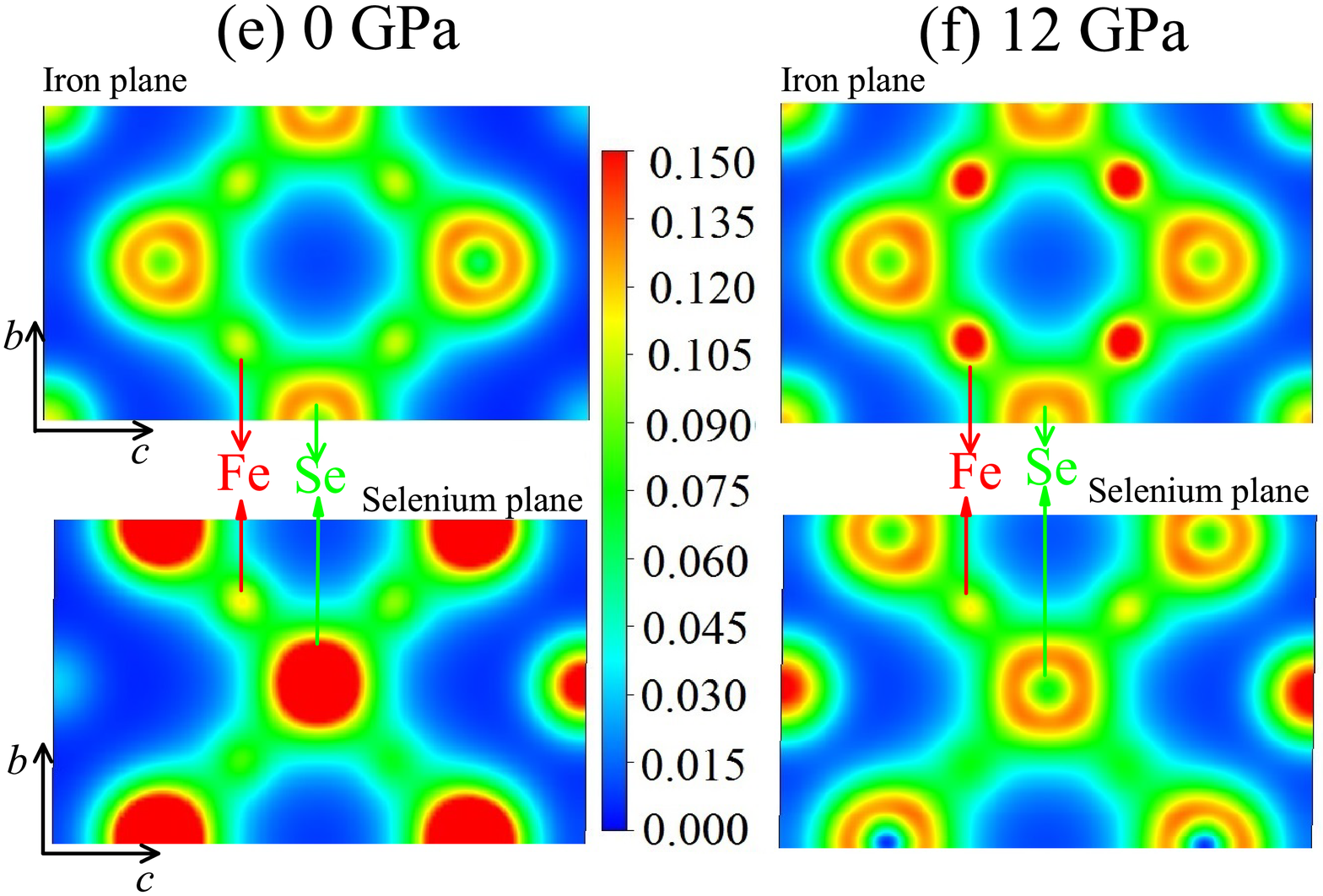}
\caption{Electronic structure of BaFe$_2$Se$_3$, from DFT.
(a-b) Band structures at $0$~GPa and $12$~GPa for the hypothetical NM state
for comparison. The Fermi level is shown with dashed lines.
The bandwidths of the iron bands increase with increasing pressure
(e.g. see the branch highlighted in red). (c-d) Band structures at $0$~GPa (Block-B) and $12$~GPa (CX)
for the magnetic states for comparison. (e-f) Electron density profiles of the iron ladders at $0$~GPa and $12$~GPa
for the hypothetical NM state for comparison.
The increase in the red intensity of the iron atoms at $12$~GPa, as compared to $0$~GPa, is indicative
of electronic doping by pressure. Upper panel: cut in an iron ladder plane; Lower panel: cut in a Se plane.}
\label{Fig5}
\end{figure}

To better understand the underlying driving force for pressure-driven phase transitions in BaFe$_2$Se$_3$,
the band structure of the NM state at $0$~GPa and $12$~GPa are displayed in Figs.~\ref{Fig5}(a-b).
This figure reveals a clear tendency for the bandwidths of the iron $3d$ bands to be
enlarged upon pressure, implying an enhancement of itinerant properties
of the $3d$ electrons.

Moreover, the band structures for magnetic ground states are displayed in Figs.~\ref{Fig5}(c-d)
at $0$ GPa and $12$ GPa, respectively. These two band structures are rather different, making it
impossible to extract more basic driving forces behind the magnetic transition. Thus, the above
discussion based on the hypothetic NM state instead of the real magnetic state is valuable to reveal
the direct pressure effects to the band structure.

The electronic density profiles of the iron ladders are
correspondingly shown in Figs.~\ref{Fig5}(e-f). It is clear that the electronic density at the iron sites
increases with increasing pressure. In fact, according to the Bader charge analysis~\cite{Bader:book,Henkelman:Cms},
electrons transfer from Se to Fe atoms by an amount $0.18$ electron/Fe
when the pressure increases from $0$~GPa to $12$~GPa.
This tendency is equivalent to doping the system with electrons: increasing pressure amounts
to adding electrons to the ladders. This same conclusion was reached in previous DFT investigations of
BaFe$_2$S$_3$ under high pressure~\cite{Zhang:prb17}. This phenomenon is compatible with recent
Density Matrix Renormalization Group (DMRG) technique results where pairing of carriers was unveiled
by doping the parent state of two-leg ladders, modeled with a two-orbital Hubbard model~\cite{Patel:prb,Patel:prb17}.
This general tendency of pressure inducing doping of ladders was previously
observed experimentally in two-leg Cu-oxide ladders as well~\cite{piskunov:prb05}. In this case hole pairing was theoretically
predicted upon doping the system and later confirmed experimentally~\cite{dagotto:science96,Dagotto:Rpp,uehara:jpsj96}.

%Besides the electronic density profiles, the electron localization function (ELF) was here also calculated
%to describe the level of electron localization~\cite{Silvi:Nat,Savini:aci}. Results are in
%Fig.~\ref{Fig5}(e-f). It is clear that the Fe-Se bond becomes more metallic for
%the CX state, while it is highly localized in the Block-B state. The ELF for NM states
%were also calculated for comparison (not shown here), and they do not show significant difference
%between $0$~GPa and $12$~GPa. Thus, the magnetic transition is the dominant driving force
%for the changes observed in the electron localization.

\section{Conclusions}

The pressure effects on BaFe$_2$Se$_3$ have been investigated using first-principles calculations.
The existence of several sequential transitions, including structural, electronic, and magnetic
transitions, have been revealed by our calculations. First, the crystalline structure transition
from $Pnma$ to $Cmcm$ occurs at $6$~GPa and it is a smooth second
order transition, which does not affect strongly any other physical properties.
Then, an insulator-metal transition happens
at $10.4$~GPa, acting as a precursor of the magnetic transition from the (unusual) Block-B antiferromagnetic state
to the (more common) CX antiferromagnetic state at $\sim 12$~GPa. Finally, all magnetic moments
quench to zero at $\sim 30$~GPa.

Our calculations have not only reproduced several experimental observations
(e.g. magnetic-nonmagnetic transition at $\sim30$ GPa and insulator-metal transition at $\sim10$ GPa)
but they have also provided a systematic
description of the phase evolution of BaFe$_2$Se$_3$
under high pressure. In addition, according to our phase diagram, we conjecture that the experimentally
observed superconducting dome could be induced by magnetic fluctuations due to the competition
between the two dominant magnetic states Block-B and CX, which is unexpected since superconducting phases in
electronic correlated systems are usually observed at the transition from a magnetic state with long-range
order to a non-magnetic state with short-range order. Moreover,  by analogy with what occurs in BaFe$_2$S$_3$
we  believe that an additional superconducting dome, potentially with an even higher $T_{\rm c}$, could exist beyond $30$~GPa
when the CX state is replaced by a non-magnetic state.
Overall, it is clear that ladder compounds in the iron-superconductors context have rich phase diagrams.
Moreover, pairing calculations in model Hamiltonians are easier in one dimension than higher, due to the availability
of powerful computational techniques. For all these reasons,
the iron ladder materials have much potential to solve the mysteries of iron-based high critical temperature superconductors.

\acknowledgments{We acknowledge valuable discussions with H.-C. Lei, K. Yamauchi, S. Picozzi, N. Patel, and Y.-F. Zhang. We also thank J.-J. Ying, V.-V. Struzhkin, and C. Petrovic for share their preprint before publication.
This work was supported by National Natural Science Foundation of China (Grant No. 11674055). L.F.L.
was supported by Jiangsu Innovation Projects for Graduate Students (Grant No. KYCX17\_0047).
E.D. was supported by the U.S. Department of Energy, Office of Basic Energy Sciences,
Materials Sciences and Engineering Division.
Most calculations were carried out at the National Supercomputer Center in Guangzhou (Tianhe-II).}

\bibliographystyle{apsrev4-1}
\bibliography{ref3}
\end{document}